\documentclass{aa}
\usepackage{graphicx}
\usepackage{txfonts}
\usepackage{amsmath}    
\usepackage{amssymb}    
\usepackage{multicol}        
\usepackage{bm}         
\usepackage{pdflscape}  
\usepackage{booktabs}
\usepackage{listings}
\usepackage{color}
\usepackage{threeparttable}
\usepackage{multirow}
\usepackage{enumerate}
\usepackage{indentfirst} 
\usepackage{stackengine}
\newcommand\xrowht[2][0]{\addstackgap[.5\dimexpr#2\relax]{\vphantom{#1}}}
\usepackage{booktabs}
\usepackage{subcaption}
\usepackage{lscape}
\usepackage{txfonts}
\usepackage{ulem}
 \usepackage[bookmarks=false,         
     pdfnewwindow=true,      
     colorlinks=true,    
     linkcolor=blue,     
     citecolor=blue,     
     filecolor=blue,  
     urlcolor=blue,      
final=true
 ]{hyperref}
 \usepackage{etoolbox}
 \usepackage{color}
\usepackage{CJK}
\usepackage[figuresright]{rotating}
\usepackage{longtable}
\usepackage{orcidlink}

\begin{document} 

\begin{CJK*}{UTF8}{gbsn}

   \title{The fundamental plane of black hole activity for low-luminosity radio active galactic nuclei across $0<z<4$\thanks{All the low-luminosity radio AGNs are summarized in Table A1 that is only available in electronic form at the CDS via anonymous ftp to \href{http://cdsarc.u-strasbg.fr}{cdsarc.u-strasbg.fr (130.79.128.5)} or via \href{http://cdsweb.u-strasbg.fr/cgi-bin/qcat?J/A+A/}{http://cdsweb.u-strasbg.fr/cgi-bin/qcat?J/A+A/}.}}

   \author{Yijun~Wang\thanks{\email{wangyijun@nju.edu.cn}} 
          \inst{1,2}
          \orcidlink{0000-0002-1010-7763},
          Tao~Wang\thanks{\email{taowang@nju.edu.cn}} 
          \inst{1,2}
          \orcidlink{0000-0002-2504-2421},
          Luis C. Ho 
          \inst{3,4}
          \orcidlink{0000-0001-6947-5846},
          Yuxing~Zhong 
          \inst{5},
          Bin~Luo 
          \inst{1,2}
          }

   \institute{School of Astronomy and Space Science, Nanjing University, 163 Xianlin Avenue, Nanjing 210023, People's Republic of China
         \and Key Laboratory of Modern Astronomy and Astrophysics, Nanjing University, Ministry of Education, 163 Xianlin Avenue, Nanjing 210023, People's Republic of China
         \and Kavli Institute for Astronomy and Astrophysics, Peking University, Beijing 100871, People's Republic of China
         \and Department of Astronomy, School of Physics, Peking University, Beijing 100871, People's Republic of China
         \and Department of Pure and Applied Physics, Waseda University, 3-4-1 Okubo, Shinjuku, Tokyo 169-8555, Japan
             }

  \abstract
 {The fundamental plane of black hole activity describes the correlation between radio luminosity ($L_{\rm R}$), X-ray luminosity ($L_{\rm X}$), and black hole mass ($M_{\rm BH}$). It reflects a connection between the accretion disc and the jet.
 However, the dependence of the fundamental plane on various physical properties of active galactic nuclei (AGNs) and host galaxies remains unclear, especially for low-luminosity AGNs, which is important for understanding the accretion physics in AGNs.}
{Here, we explore the dependence of the fundamental plane on the radio loudness, Eddington-ratio ($\lambda_{\rm Edd}$), redshift, and galaxy star formation properties (star-forming galaxies and quiescent galaxies) across $0.1 < z \leq 4$ for radio AGNs. Based on current deep and large surveys, our studies can extend to lower luminosities and higher redshifts.} 
{From the deep and large multi-wavelength surveys in the GOODS-N, GOODS-S, and COSMOS/UltraVISTA fields, we constructed a large and homogeneous radio AGN sample consisting of 208 objects with available estimates for $L_{\rm R}$ and $L_{\rm X}$. Then we divided the radio AGN sample into 141 radio-quiet AGNs and 67 radio-loud AGNs according to the radio loudness defined by the ratio of $L_{\rm R}$ to $L_{\rm X}$, and explored the dependence of the fundamental plane on different physical properties of the two populations, separately.
}
{The ratio of $L_{\rm R}$ to $L_{\rm X}$ shows a bimodal distribution that is well described by two single Gaussian models. The cross point between these two Gaussian components corresponds to a radio-loudness threshold of $\log (L_{\rm R}/L_{\rm X}) = -2.73$. The radio-quiet AGNs have a significantly larger Eddington ratio than the radio-loud AGNs. Our radio-quiet and radio-loud AGNs show a significantly different fundamental plane, which indicates a significant dependence of the fundamental plane on the radio loudness.  
For both radio-quiet and radio-loud AGNs, the fundamental plane shows a significant dependence on $\lambda_{\rm Edd}$, but no dependence on redshift. The fundamental plane shows a significant dependence on the galaxy star formation properties for radio-quiet AGNs, while for radio-loud AGNs this dependence disappears.
}
{The fundamental plane sheds important light on the accretion physics and X-ray emission origins of central engines.
X-ray emission of radio-quiet AGNs at $0.01 < \lambda_{\rm Edd} < 0.1$ are produced by a combination of advection-dominated accretion flow (ADAF) and synchrotron radiation from the jet, while at $0.1 < \lambda_{\rm Edd} < 1$ they mainly follow the synchrotron jet model. The origins of X-ray emission of radio-loud AGNs are consistent with a combination of ADAF and the synchrotron jet model at $\lambda_{\rm Edd} < 0.01$, agree with the synchrotron jet model at $0.01 < \lambda_{\rm Edd} < 0.1$, and follow a combination of the standard thin disc and a jet model at $\lambda_{\rm Edd} > 0.1$.
}


   \authorrunning{Y.J. Wang et al.}
   \titlerunning{the fundamental plane of black hole activity for low-luminosity radio AGNs at $0 < z < 4$}
   \maketitle

%
%
\section{Introduction}
\label{sec:intro}

Accreting supermassive black holes (SMBHs), also known as active galactic nuclei (AGNs), emit
immense energy across the whole electromagnetic spectrum, which is believed to have a great impact on the growth and evolution of host galaxies \citep[][for reviews]{Fabian2012,King2015}. Thus, AGNs are ideal laboratories in which to explore both accretion physics around black holes and their connection with host galaxies. 
Accretion physics around black holes are found to be scale-invariant
across black hole mass scale from $\sim 10$ solar masses of X-ray binaries (XRBs) to $10^6 \sim 10^{10}$ solar masses of SMBHs \citep{Merloni2003,Falcke2004,Done2005,McHardy2006,Kording2014,Ruan2019}.
One of the most prominent pieces of evidence supporting the unification of XRBs and SMBHs is the fundamental plane of black hole activity \citep[][and references therein]{Merloni2003,Falcke2004} 
that is characterized by a nonlinear empirical relation given by radio luminosity, X-ray luminosity, and black hole mass.
The radio luminosity is thought to be related to jet activities \citep{Begelman1984}, while the ratio of X-ray luminosity to black hole mass is usually taken as a tracer for the accretion rate of the disc \citep{Haardt1991,Liu2022}.
Thus, the fundamental plane connecting XRBs and SMBHs suggests a similar disc-jet connection across different mass scales \citep[e.g.,][]{Merloni2003,Falcke2004,Plotkin2012,Dong2014}. The fundamental plane also provides an approach to estimating black hole mass directly through radio luminosity and X-ray luminosity \citep[e.g.,][and references therein]{Merloni2003,Gultekin2019}.

However, a growing number of studies focusing only on the fundamental plane of AGNs demonstrate that different types of AGN, such as low-ionization nuclear emission line regions (LINERS), Seyferts, and quasars, show different fundamental planes \citep[e.g.,][]{Yuan2009,Gultekin2009,Bonchi2013,Saikia2015,Nisbet2016,Fan2016,Xie2017,Li2018,Liao2020,Bariuan2022}. These results indicate that the fundamental plane may depend on the accretion state of the disc or be sensitive to the adopted sample of black holes \citep{Plotkin2012}.
In order to increase sample sizes, the majority of the above-mentioned studies about the fundamental plane used a combination of radio-quiet and radio-loud AGNs. For the first time, \cite{Wang2006} and \cite{Li2008} studied the fundamental plane in broad-line radio-quiet and radio-loud AGNs separately. More recently, \cite{Bariuan2022} focused on radio-quiet and radio-loud quasars. These works all found that radio-quiet and radio-loud AGNs show quite different fundamental planes and follow different theoretical accretion models given by \cite{Merloni2003}.
Due to limitations of data, these works mainly focus on high-luminosity radio-loud and radio-quiet AGNs. Thanks to the deep and large multi-wavelength surveys in the GOODS-N, GOODS-S, and COSMOS/UltraVISTA fields \citep[e.g.,][]{Owen2018,Alberts2020,Smolcic2017a,Liu2018,Jin2018}, we can extend the fundamental plane studies to low-luminosity radio-quiet and radio-loud AGNs.

Due to the limitation of sample sizes, in the past it was difficult to analyse in detail the dependence of the fundamental plane on other physical properties, such as 1) the Eddington ratio, 2) redshift, and 3) galaxy star formation properties, especially for low-luminosity AGNs. These studies may be crucial to our understanding of why different types of AGNs exhibit different fundamental planes. 
1) The ``Eddington ratio:'' Observational evidences show that radio-quiet and radio-loud AGNs have different fundamental planes \citep{Wang2006,Li2008,Bariuan2022}, and radio loudness exhibits a negative correlation with Eddington ratio \citep{Ho2002,Panessa2007,Sikora2007}. This demonstrates that the fundamental plane may depend on the Eddington-scaled accretion rate, which is still lacking in quantitative studies until now. 
2) ``Redshift:'' The AGN accretion rate density peaks at $z \sim$ 2, and declines toward the local Universe \citep[see][for a review]{Aird2010,Madau2014}, which may give rise to different accretion physics at different cosmic times. The majority of previous works studied the fundamental plane at low redshift ($z<0.5$), while some of them extended to high redshift but only for high-luminosity AGNs \citep{Wang2006,Li2008,Bariuan2022}. For low-luminosity AGNs, the dependence of the fundamental plane on redshift is still unclear.
3) ``Galaxy star formation properties:'' Radio AGNs hosted by star-forming galaxies (SFGs) and quiescent galaxies (QGs) have different cosmic evolutions for the AGN incident rate \citep[e.g.,][hereafter ``Paper I'']{Janssen2012,Kondapally2022,Wang2024} and radio luminosity functions \citep[][Paper I]{Kondapally2022}. Given that SFGs and QGs may perform different fueling mechanisms toward central SMBHs \citep[e.g.,][]{Kauffmann2009,Kondapally2022,Ni2023},
radio activities of central engines may depend on the fueling mechanisms of their host galaxies. Further, it still remains unclear whether the fueling mechanisms affect the disc-jet connection characterized by the fundamental plane.

In this work, we first introduce a parent radio AGN sample from the GOODS-N, the GOODS-S, and the COSMOS/UltraVISTA fields selected by Paper I across $0.1 < z \leq 4$ (Section \ref{sec:parentAGN}). For these radio AGNs, we collected available measurements for radio luminosity and X-ray luminosity from our previous works and other literature works, and inferred black hole mass from stellar mass (Section \ref{sec:physicalpara}). Next, we divided our sample into radio-quiet and radio-loud AGN subsamples according to the radio loudness defined by the relative strength of the radio and X-ray emission (Section \ref{sec:RQRLsample}). Then, we fit the fundamental plane for the radio-quiet and radio-loud AGN subsamples separately (Section \ref{sec:fitness}). Further, we give a brief summary about the fundamental plane studies and discuss the dependence of the fundamental plane on Eddington ratio, redshift, and galaxy star formation properties (Section \ref{sec:discussion}). In Section \ref{sec:discussion}, we also discuss the central engines for radio-quiet and radio-loud AGNs. 
Finally, we summarize our conclusions in Section \ref{sec:summary}.
Throughout this paper, we assume a \cite{Chabrier2003} initial mass function (IMF)
and a flat cosmology with the following parameters:
$\Omega_{\rm m}=0.3$, $\Omega_{\Lambda}=0.7$, and $H_0=70\ \rm{km}\ \rm{s}^{-1}\ \rm{Mpc}^{-1}$.

\section{Parent radio AGN sample}
\label{sec:parentAGN}

\subsection{GOODS-N and COSMOS fields}
\label{sec:GNCS}
For the GOODS-N and COSMOS/UltraVISTA fields, our parent radio AGN sample was derived from Paper I, which consists of 102 radio AGNs from the GOODS-N and 881 radio AGNs from the COSMOS/UltraVISTA across $0.1 < z <4$ selected by the infrared-radio correlation \citep[IRRC; e.g., ][and references therein]{Helou1985,Condon1992}.
We refer to Section 4 in Paper I for more details about sample selections and here we just give a brief introduction.
The IR emission and radio emission of the star formation process have a mutual origin in the activities of massive stars \citep{Condon1992,Dubner2015}, which results in a tight correlation between them (IRRC). 
The IRRC is usually defined by the ratio ($q_{\rm TIR}$) of rest-frame 8--1000 $\mu$m IR luminosity ($L_{\rm{TIR}}$) 
to rest-frame 1.4 GHz radio luminosity ($L_{1.4 \rm{GHz}}$),
which is in the form of $q_{\rm{TIR}} = \log [ L_{\rm{TIR}}/(L_{1.4 \rm{GHz}} \times 3.75 \times 10^{12}\ \rm{Hz}) ]$ \citep{Helou1985}.
The AGNs may have extra radio emission from nuclear activities, such as jets \citep[][for a review]{Panessa2019}. 
Therefore, radio AGNs usually exhibit a radio excess relative to the IRRC and have a smaller $q_{\rm{TIR}}$ value than that of star formation.
In Paper I, we define a $q_{\rm{TIR}}$ threshold ($q_{\rm{TIR,AGN}}$) to select radio AGNs (see details in Paper I).
A radio source will be selected as a radio AGN if its $q_{\rm{TIR}}$ value is lower than $q_{\rm{TIR,AGN}}$.
In addition, 
$L_{1.4 \rm{GHz}}$ was derived from Paper I and was calculated by the radio flux from the deep VLA 1.4 GHz \citep[GOODS-N; ][]{Owen2018} or 3 GHz radio surveys \citep[COSMOS; ][]{Smolcic2017a} assuming a radio spectral index of $-0.8$ \citep[e.g.,][]{Yang2022}.
$L_{\rm TIR}$ was derived from Paper I and was estimated by the broadband spectral energy distribution (SED) fitting.
In the SED fitting process, the far-infrared (FIR) and submillimeter data are derived from the ``super de-blended'' photometry \cite[][]{Liu2018,Jin2018}, ensuring a more precise estimate for the IR luminosity.

\subsection{GOODS-S field}
\label{sec:GS}
We also selected a radio AGN sample from the GOODS-S field following Paper I to enlarge our sample.
We first derived radio data at 3 GHz from \cite{Alberts2020}, which contains 712 sources with a signal-to-noise ratio (S/N) larger than 3.
Then we crossmatched this 3 GHz catalog with the ultraviolet-optical-mid-infrared (UV-optical-MIR) catalog from \cite{Guo2013} by a match radius of 1 arcsec.
This UV-optical-MIR catalog covers the 
wavelength range between 0.4 and 8 $\mu$m, and contains 
34,930 sources over 171 arcmin$^2$.
After crossmatching, 414 (out of 712) radio objects in \cite{Alberts2020} have UV/optical/MIR counterparts in \cite{Guo2013}.
Following the selection criteria in Paper I ($0.1 < z \leq 4.0$ and S/N of radio flux $\geq 5$), we selected 366 radio sources from the 414 objects as our radio source sample in the GOODS-S field to conduct further analysis.
Next, we collected MIR-FIR data in the GOODS-S field from Wang et al. (in prep.).
Based on the similar ``super de-blended'' photometry method in the GOODS-N and COSMOS fields,
Wang et al. (in prep.) obtained photometry in the
MIR band ({\it Spitzer} 16 and 24 $\mu$m) and ``super de-blended'' photometry 
in the FIR band ({\it Herschel} 100, 160, 250, 350, and 500 $\mu$m) for 1881 objects.
We also utilized the submillimeter data derived from the
SCUBA-2 850 $\mu$m survey \citep{Cowie2018}, 
the Atacama Large Millimeter/submillimeter Array (ALMA) 870 $\mu$m survey \citep{Tadaki2020}, 
and the ALMA 1.1 mm survey \citep{Gmez-Guijarro2022}. 
Further, we crossmatched the radio source sample in the GOODS-S field
with the MIR-FIR catalog (Wang et al. in prep.) by a match radius of 1 arcsec,
with the SCUBA-2 850 $\mu$m survey 
by a radius of 5 arcsec,
with the ALMA 870 $\mu$m survey  
by a radius of 1 arcsec,
and with the ALMA 1.1 mm survey  
by a radius of 1.5 arcsec.
For the 366 sources in the radio source sample, 324 objects have MIR-FIR counterparts in Wang et al. (in prep.).
Of these, 27 (out of 324) objects have SCUBA-2 850 $\mu$m detections,
12 (out of 324) objects have ALMA 870 $\mu$m detections,
and 35 (out of 324) objects have ALMA 1.1 mm detections.
For the above-mentioned 324 objects that have multi-wavelength data 
from the UV-optical-MIR to FIR-submillimeter-radio bands,
we performed a broadband SED fitting with Code Investigating GALaxy Emission 
\citep[\textsc{cigale} 2022.0;][]{Burgarella2005,Noll2009,Boquien2019,Yang2020,Yang2022} to 
estimate various galaxy properties 
(such as $L_{\rm TIR}$ and 
the stellar mass, $M_{\star}$).
We refer to Appendix A in Paper I for parameter settings about the SED fitting.
Finally, we followed Paper I to select a parent radio AGN sample in the GOODS-S field including 71 sources across $0.1 < z \leq 4.0$ (see brief introduction in Section \ref{sec:GNCS} and detailed analysis in Section 4 of Paper I).

\section{Derived quantities}
\label{sec:physicalpara}

\subsection{Black hole mass ($M_{\rm BH}$)}
In this work, we estimated the black hole mass in an indirect way that is based on the correlation between the black hole mass and the total stellar mass \citep[$M_{\rm BH}$-$M_\star$ relation; e.g.,][for a review]{Greene2020}.
Here, $M_\star$ was estimated by the broadband SED fitting
(see details in Paper I for the GOODS-N and COSMOS fields, and see details in Section \ref{sec:GS} for the GOODS-S field). 
Most of the radio AGNs in our sample are hosted by ellipticals.
Thus, we used the $M_{\rm BH}$-$M_\star$ relation in the form of 
$\log (M_{\rm BH}/M_\odot)=(7.89\pm 0.09)+(1.33\pm 0.12)\times \log (M_\star/3\times10^{10}M_\odot)$
from \cite{Greene2020} for local early-type galaxies.
The $M_\star$ in this work was estimated based on a \cite{Chabrier2003} IMF, 
while \cite{Greene2020} adopted a diet Salpeter IMF \citep{Bell2003a}.
Thus, we divided our $M_\star$ by a constant value of 1.14 \citep{Madau2014,vanderWel2007} to convert values from the \cite{Chabrier2003} IMF to the diet Salpeter IMF.
In addition, \cite{Li2023} found that the $M_\star$-$M_{\rm BH}$ relation does not show a significant evolution with redshift across $0 < z < 3.5$.
Therefore, we applied the local $M_\star$-$M_{\rm BH}$ relation from \cite{Greene2020} to the whole redshift range in this work ($0.1 < z < 4$).
The uncertainties of $M_{\rm BH}$ were inferred from the parameter errors of the $M_{\rm BH}$-$M_\star$ relation \citep{Greene2020} using the propagation of error method.

\begin{figure*}
\centering
\includegraphics[width=\linewidth, clip]{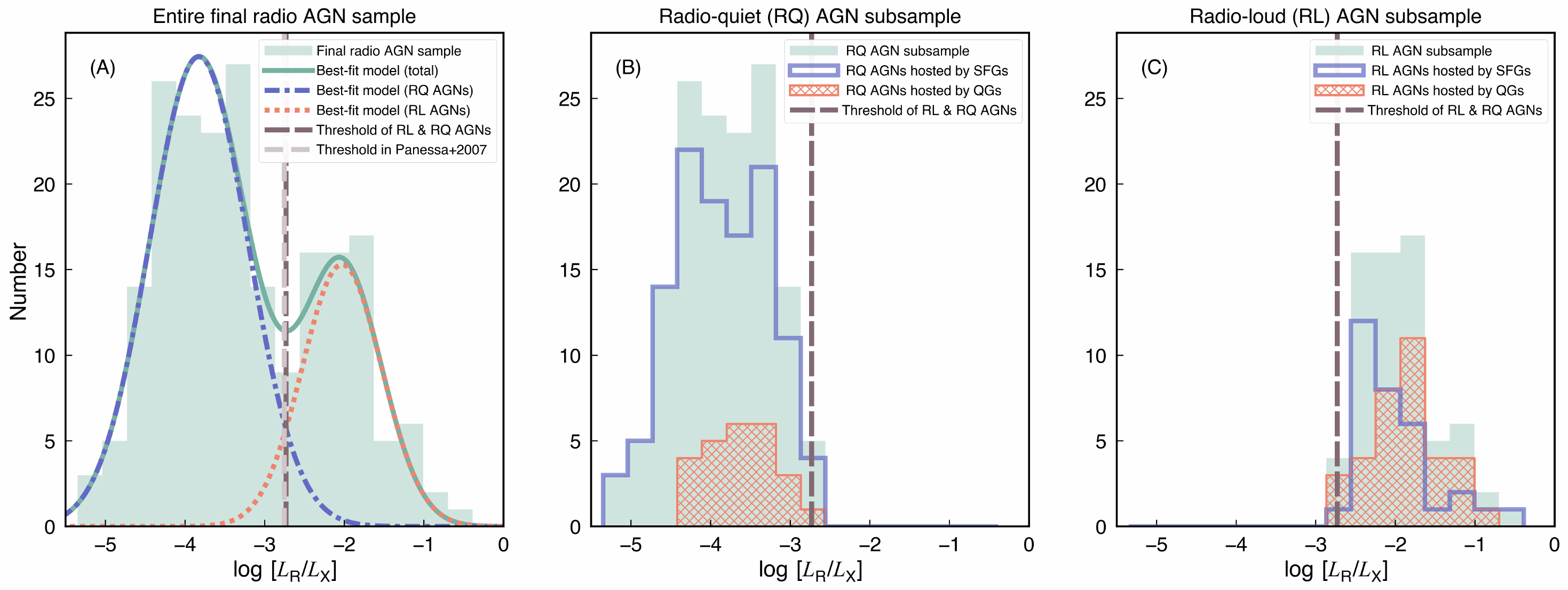}
\caption{Distribution of the ratio of the rest-frame 5 GHz radio luminosity ($L_{\rm R}$) to the rest-frame 2--10 keV X-ray luminosity ($L_{\rm X}$) for the entire final radio AGN sample (Panel A), the radio-quiet AGN subsample (Panel B), and the radio-loud AGN subsample (Panel C). The solid green curve in Panel (A) represents the best-fit model to the entire distribution, which consists of two single Gaussian models (the dash-dotted blue curve and the dotted red curve). The vertical dashed dark gray line in all panels represents the cross point between two single models, which is defined as the radio-loudness threshold to divide radio-loud and radio-quiet AGNs in this work. The vertical dashed light gray line in Panel (A) represents the radio-loudness threshold in \cite{Panessa2007}, which was obtained by a sample consisting of local Seyferts and low-luminosity radio galaxies. The blue and red histograms in Panel (B) represent the radio-quiet AGNs hosted by SFGs and QGs, respectively, while in Panel (C), they represent the radio-loud AGNs hosted by SFGs and QGs, respectively.
\label{fig:LRLXdis}}
\end{figure*}

\subsection{Rest-frame 5 GHz radio luminosity ($L_{\rm R}$)}
To get a pure radio luminosity for AGNs, radio luminosity from star formation ($L_{\rm{SF}}$) should be subtracted from the total radio luminosity.
$L_{\rm{SF}}$ was estimated by the IRRC in the form of 
\begin{equation}
L_{1.4 \rm{GHz}, \rm{SF}} = L_{\rm{TIR}}/(3.75 \times 10^{12}\ {\rm Hz} \times 10^{q_{\rm{TIR}}})
\label{eq:LRSF}
\end{equation}
\citep{Helou1985}, where $L_{\rm TIR}$ is the rest-frame 8--1000 $\mu$m IR luminosity, 
$L_{1.4 \rm{GHz}, \rm{SF}}$ is the rest-frame 1.4 GHz radio luminosity from star formation in units of erg s$^{-1}$ Hz$^{-1}$,
and $q_{\rm{TIR}}$ is the IRRC index for the star formation. $q_{\rm{TIR}}$ has been found to depend on redshift \citep{Magnelli2015,Delhaize2017,Novak2017,Enia2022} and stellar mass \citep{Delvecchio2022}. For consistency, here we used the relation obtained in our Paper I, which is in the form of $q_{\rm{TIR}} = (2.62\pm 0.08)\times (1+z)^{-0.08\pm 0.03}$. In Paper I, we did not consider a $M_{\star}$-dependent IRRC as it does not have a significant impact on our results.

Then the rest-frame 5 GHz radio luminosity from AGN (hereafter $L_{\rm R}$, in units of erg s$^{-1}$) was calculated by
\begin{equation}
L_{\rm R}=\frac{4\pi D_L^2}{(1+z)^{1+\alpha_{\rm AGN}}}\times \left(\frac{5\ {\rm GHz}}{\nu_{\rm obs}} \right)^{\alpha_{\rm AGN}}\times (S_{\nu, \rm{obs}}-S_{\nu, \rm{SF}})\times 5\ {\rm GHz}.
\label{equ:LRequ}
\end{equation}
Here, $D_L$ is the luminosity distance (in centimeters), $z$ is the redshift, 
$\nu_{\rm obs}$ is the observed frequency (in GHz),
and $S_{\nu, \rm{obs}}$ (in units of erg s$^{-1}$ cm$^{-2}$ Hz$^{-1}$) is the observed integrated flux densities at $\nu_{\rm obs}$.
For GOODS-N, $S_{\nu, \rm{obs}}$ at $\nu_{\rm obs}=1.4$ GHz was derived from \cite{Owen2018} (see Section \ref{sec:GNCS} or Paper I).
For GOODS-S, $S_{\nu, \rm{obs}}$ at $\nu_{\rm obs}=3$ GHz was derived from \cite{Alberts2020} (see Section \ref{sec:GS}).
For COSMOS, $S_{\nu, \rm{obs}}$ at $\nu_{\rm obs}=3$ GHz was derived from \cite{Smolcic2017a} (see Section \ref{sec:GNCS} or Paper I).
$S_{\nu, \rm{SF}}$ in Equation \ref{equ:LRequ} is the flux densities at $\nu_{\rm obs}$ attributed to the star formation, which was calculated by 
\begin{equation}
S_{\nu, \rm{SF}}=L_{1.4 \rm{GHz}, \rm{SF}}\ \frac{(1+z)^{1+\alpha_{\rm SF}}}{4\pi D_L^2}\ \left(\frac{\nu_{\rm obs}}{1.4\ {\rm GHz}}\right)^{\alpha_{\rm SF}}
\end{equation}
\citep{Condon1992}, where $L_{1.4 \rm{GHz}, \rm{SF}}$ is given by Equation \ref{eq:LRSF} and $\alpha_{\rm SF}$ is the radio spectral index for star formation, which is assumed to be $-0.8$ \citep[e.g.,][]{Yang2022}.
$\alpha_{\rm AGN}$ in Equation \ref{equ:LRequ} is the radio spectral index for AGNs,
which is assumed to be $-0.5$ \citep[e.g.,][]{Kellermann1989,Gasperin2018} throughout this work.
The value of $\alpha_{\rm AGN}$ is usually related to the radio morphology (e.g., $\alpha_{\rm AGN} \sim 0.0$ for core-dominated radio AGNs), which may change from object to object.
Therefore, we also calculated $L_{\rm R}$ under the assumption of $\alpha_{\rm AGN}=0.0$, $-0.1$, $-0.3$, $-0.7$, respectively,
and used the Kolmogorov-Smirnov (K-S) test to check whether different $\alpha_{\rm AGN}$ values result in different $L_{\rm R}$ distributions.
The $p$ values of all the K-S tests are much higher than 0.05,
which indicates that adopting an $\alpha_{\rm AGN}$ value (between 0.0 and $-0.7$) does not affect the calculation for $L_{\rm R}$ in our sample.
Based on the propagation of error method, the uncertainties of $L_{\rm R}$ were inferred from the flux uncertainties given by the above-mentioned radio surveys and the average measured uncertainties of $q_{\rm TIR}$ ($\sim 0.18$; see our Paper I).

In some literature works, the rest-frame 5 GHz radio luminosity in the fundamental plane refers to the nuclear radio emission from a compact radio core in order to compare with X-ray emission that are mainly from central engine. However, VLA 1.4 GHz or 3 GHz surveys in the GOODS-N, GOODS-S, and COSMOS fields do not have a high-enough spatial resolution to resolve the compact radio core. Even so, their spatial resolutions still ensure that the radio flux is measured on the top of the galaxy rather than the large-scale jet.
In addition, nearly 20\% of our sample have radio detections with a higher spatial resolution, such as VLA 10 GHz Pilot Survey in the GOODS-N field \citep[average spatial resolution of $0.22'' \approx 1.76$ kpc at $z = 1$;][]{Murphy2017} and VLA 6 GHz Survey in the GOODS-S field \citep[average angular resolution of $0.62'' \times 0.31'' \approx 4.97\ {\rm kpc} \times 2.48$ kpc at $z = 1$;][]{Alberts2020,Lyu2022}. These objects do not show significant large-scale jet structures.
Furthermore, we utilized the ratio of peak radio flux to total radio flux as a proxy for compactness to select a sample only including objects with a relatively compact structure (including 110 sources). We found that using this compact object subsample does not change our results compared to using the entire sample. In order to ensure a large sample to conduct the analysis, in this work we used the entire sample and have not considered whether the rest-frame 5 GHz radio luminosity is strictly derived from a spatially resolved compact core.

\begin{figure*}
\includegraphics[width=\linewidth, clip]{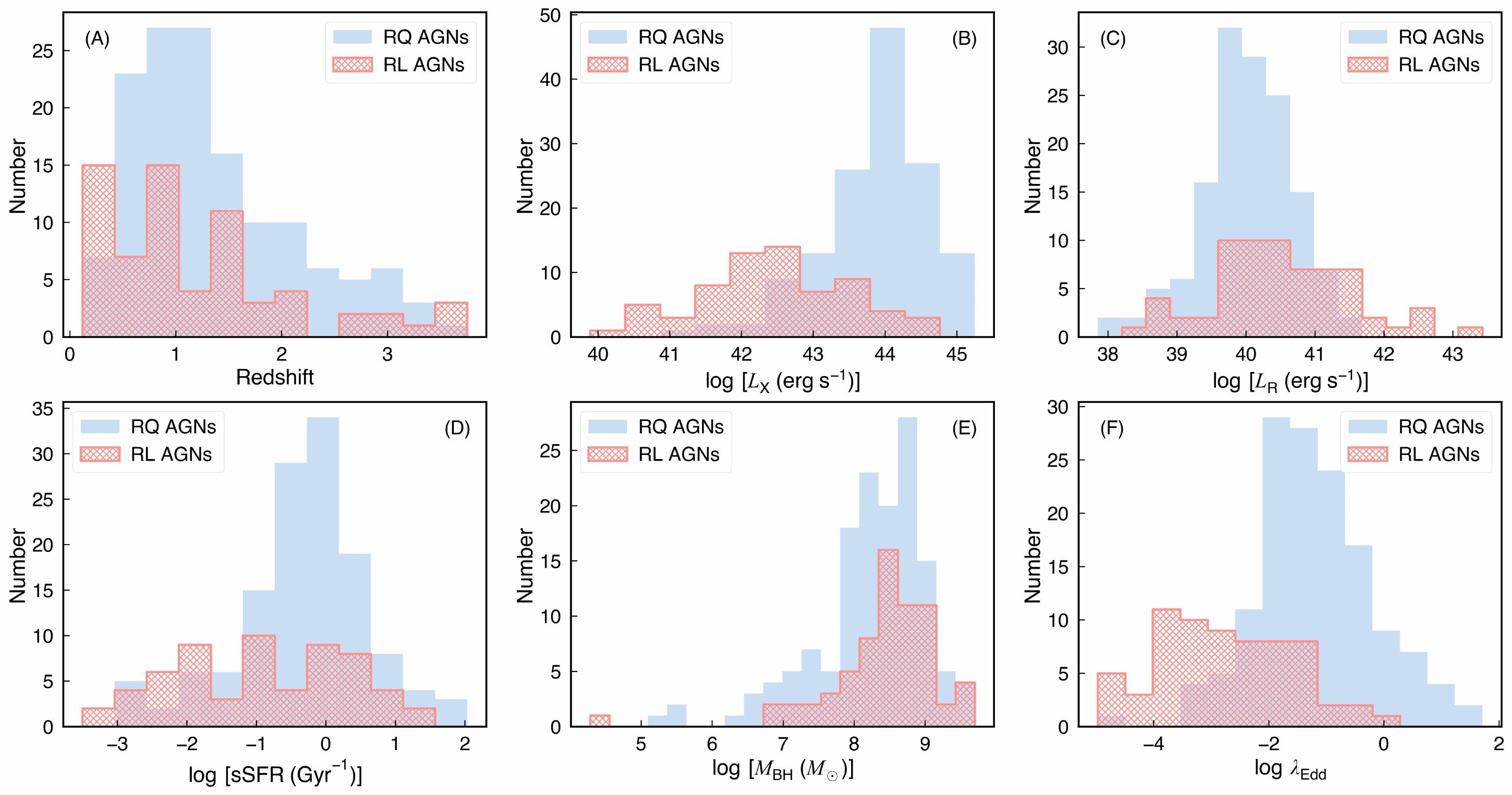}
\caption{Distribution of physical properties for the radio-quiet (RQ) AGNs (blue histogram) and radio-loud (RL) AGNs (red histogram) used to study the fundamental plane of black hole activity, including: (A) redshift, (B) rest-frame 2--10 keV X-ray luminosity from AGNs ($L_{\rm X}$), (C) rest-frame 5 GHz radio luminosity from AGNs ($L_{\rm R}$), (D) the specific star formation rate (sSFR), (E) the black hole mass ($M_{\rm BH}$) inferred from stellar mass, and
(F) the Eddington ratio ($\lambda_{\rm Edd}$).
\label{fig:RLRQproperties}}
\end{figure*}

\subsection{Rest-frame 2--10 keV X-ray luminosity ($L_{\rm X}$)}
\label{sec:LXpara}
X-ray data in the GOODS-N were mainly derived from the 2 Ms {\it Chandra} Deep Field-North
(CDF-N) survey \citep{Xue2016}.
\cite{Xue2016} estimated the absorption-corrected rest-frame 0.5--7 keV luminosity assuming an intrinsic photon index of $\Gamma=1.8$ \citep[typical value for the X-ray spectrum of AGNs;][]{Tozzi2006}.
Then we crossmatched our parent radio AGN sample in the GOODS-N (102 objects, see Section \ref{sec:GNCS}) with the 2 Ms CDF-N survey by a match radius of 1.5 arcsec.
A total of 46 (out of 102) objects have X-ray counterparts and their intrinsic 2--10 keV luminosities used in this work were calculated by the 
intrinsic 0.5--7 keV luminosity from \cite{Xue2016} under the assumption of $\Gamma=1.8$. Among these 46 objects, 17 sources underwent systematic X-ray spectral analysis in \cite{Li2019} ($\Gamma$ is a free parameter in the fit and $\Gamma = 1.80\pm 0.08$ for these objects). Therefore, the intrinsic 2--10 keV luminosities of these 17 sources were derived from \cite{Li2019}.

X-ray data in the GOODS-S were mainly derived from the 7 Ms {\it Chandra} Deep Field-South
(CDF-S) survey \citep[][]{Luo2017}. Similar to the GOODS-N, the absorption-corrected rest-frame 0.5--7 keV luminosities in the GOODS-S were estimated by assuming an intrinsic photon index of $\Gamma=1.8$ \citep[][]{Luo2017}.
We crossmatched our parent radio AGN sample in the GOODS-S (71 objects, see Section \ref{sec:GS}) with the 7 Ms CDF-S survey by a match radius of 1.5 arcsec.
A total of 48 (out of 71) objects have X-ray counterparts and their intrinsic 2--10 keV luminosities used in this work were calculated by the 
intrinsic 0.5--7 keV luminosity from \cite{Luo2017} under the assumption of $\Gamma=1.8$.
Among these 48 objects, 23 sources underwent systematic X-ray spectral analysis in \cite{Liu2017} ($\Gamma$ is a free in the fit and $\Gamma = 1.81\pm 0.09$). Therefore, the intrinsic 2--10 keV luminosities of these 23 sources were derived from \cite{Liu2017}.

X-ray data in COSMOS were mainly derived from the X-ray spectral fitting for the 4.6 Ms {\it Chandra} COSMOS-Legacy survey \citep[][]{Marchesi2016} that gives the intrinsic 2--10 keV luminosity. We crossmatched our parent radio AGN sample in the COSMOS/UltraVISTA (881 objects, see Section \ref{sec:GNCS}) with the 4.6 Ms {\it Chandra} COSMOS-Legacy survey by a match radius of 1.5 arcsec. A total of 117 (out of 881) objects have X-ray counterparts with available estimates for the intrinsic 2--10 keV luminosity ($\Gamma = 1.85 \pm 0.62$). Among these 117 objects, eight are classified as Compton-thick AGNs by \cite{Lanzuisi2018}.
For these eight objects, we used their obscuration-corrected intrinsic 2--10 keV luminosity from \cite{Lanzuisi2018} to conduct the subsequent analysis.

The above-mentioned 2--10 keV luminosity refers to the total X-ray luminosity from the entire galaxy ($L_{\rm 2-10keV, tot}$). 
In order to get the pure X-ray luminosity from AGNs, X-ray radiation from XRBs should be subtracted. 
X-ray luminosities of XRBs have been found to correlate with the star formation rate (SFR) and stellar mass \citep{Grimm2003,Lehmer2010,Mineo2012,Mineo2014},
which also shows an evolution with redshift \citep{Lehmer2016}.
Here, we used the relation of 
\begin{equation}
\frac{L_{\rm 2-10keV, XRBs}}{{\rm erg\ s^{-1}}} = 10^{29.3}(1+z)^{2.19} \frac{M_{\star}}{M_{\odot}} + 10^{39.4}(1+z)^{1.02} \frac{{\rm SFR}}{M_\odot\ {\rm yr}^{-1}}
\label{eq:XRBs}
\end{equation}
derived from \cite{Lehmer2016} based on the 6Ms CDF-S survey.
The $M_{\star}$ and SFR in this relation were estimated assuming a \cite{Kroupa2001} IMF, 
while we used the \cite{Chabrier2003} IMF throughout this work.
Therefore, we first divided our $M_{\star}$ and SFR by a constant factor of 1.06 \citep{Speagle2014} to convert values from the \cite{Chabrier2003} IMF to the \cite{Kroupa2001} IMF.
Then we applied the scaled $M_{\star}$ and SFR into Equation \ref{eq:XRBs} to get the X-ray luminosity from XRBs.
Finally, the pure X-ray luminosity from AGNs (hereafter $L_{\rm X}$, in units of erg s$^{-1}$) can be calculated by
\begin{equation}
L_{\rm X} = L_{\rm 2-10keV, tot} - L_{\rm 2-10keV, XRBs},
\label{eq:LXAGN}
\end{equation}
where $L_{\rm 2-10keV, XRBs}$ is calculated by Equation \ref{eq:XRBs}.
Based on the propagation of error method, the uncertainties of $L_{\rm X}$ were inferred from the flux uncertainties given by the above-mentioned X-rays surveys and the parameter uncertainties of Equation \ref{eq:XRBs} given by \cite{Lehmer2016}.

Considering the available estimates for $M_\star$, $L_{\rm R}$, and $L_{\rm X}$ ($L_{\rm X}$ in Equation \ref{eq:LXAGN} is required to be greater than 0), 
we selected
a radio AGN sample including 208 objects (46 objects in GOODS-N, 45 in GOODS-S, and 117 in COSMOS/UltraVISTA) that were utilized in the subsequent analysis for the fundamental plane. The basic information for these 208 objects is summarized in Table \ref{tab:fitness}.
In addition, for the radio AGNs without X-ray counterparts in the aforementioned X-ray surveys, we estimated their 2--10 keV X-ray luminosity upper limits by the X-ray detection limit in these three fields \citep{Xue2016,Luo2017,Civano2016} assuming $\Gamma = 1.8$.
The X-ray detection limit is $2.0\times 10^{-16}$ erg cm$^{-2}$ s$^{-1}$ in the 0.5--7 keV band for GOODS-N \citep{Xue2016},
$1.9\times 10^{-17}$ erg cm$^{-2}$ s$^{-1}$ in the 0.5--7 keV band for GOODS-S \citep{Luo2017},
and $1.5\times 10^{-15}$ erg cm$^{-2}$ s$^{-1}$ in the 2--10 keV band for COSMOS \citep{Civano2016}.

\section{Final sample of radio-loud and radio-quiet active galactic nuclei}
\label{sec:RQRLsample}

In this work, our final radio AGN sample used to make the fundamental plane analysis
consists of the above-mentioned 208 objects, which have available estimates for $L_{\rm R}$, $L_{\rm X}$, and inferred $M_{\rm BH}$ from $M_{\star}$.
We utilized the radio loudness defined by the relative strength of the radio and X-ray emission \citep[$R_{\rm X}$; e.g.,][]{Terashima2003,Panessa2007,Ho2008} to divide radio-loud and radio-quiet AGNs in this work.
The ratio of $L_{\rm R}$ to $L_{\rm X}$ is shown in Panel (A) of Fig. \ref{fig:LRLXdis}. The $\log (L_{\rm R}/L_{\rm X})$ distribution shows a bimodal shape that is attributed to radio-quiet and radio-loud AGNs, respectively. Next, we used two single Gaussian components
to model the double peaks (see the dash-dotted blue line and dotted red line in Panel A of Fig. \ref{fig:LRLXdis}),
and used their cross point as the radio-loudness threshold (see the vertical dashed dark gray line in Panel A of Fig. \ref{fig:LRLXdis}) to divide radio-quiet and radio-loud AGNs in this work. This radio-loudness threshold ($R_{\rm X,T}$) is 
\begin{equation}
R_{\rm X,T} = \log (L_{\rm R}/L_{\rm X})_{\rm cross\ point} = -2.73. 
\label{eq:radioloudness}
\end{equation}
It is consistent with the result in \cite{Panessa2007} that was obtained based on a sample of 47 local Seyferts and 16 local low-luminosity radio galaxies (see the vertical dashed light gray line in Panel A of Fig. \ref{fig:LRLXdis}).
Finally, we selected 141 radio-quiet AGNs (68\% of the final radio AGN sample) and 67 radio-loud AGNs (32\% of the final radio AGN sample). The physical properties of these radio-quiet and radio-loud AGNs are shown in Fig. \ref{fig:RLRQproperties}.
All the radio AGNs including their various physical properties are summarized in Table A1\footnote{Table A1 is only available in electronic form at the CDS via anonymous ftp to \href{http://cdsarc.u-strasbg.fr}{cdsarc.u-strasbg.fr (130.79.128.5)} or via \href{http://cdsweb.u-strasbg.fr/cgi-bin/qcat?J/A+A/}{http://cdsweb.u-strasbg.fr/cgi-bin/qcat?J/A+A/}.}.

\begin{figure*}
\includegraphics[width=\linewidth, clip]{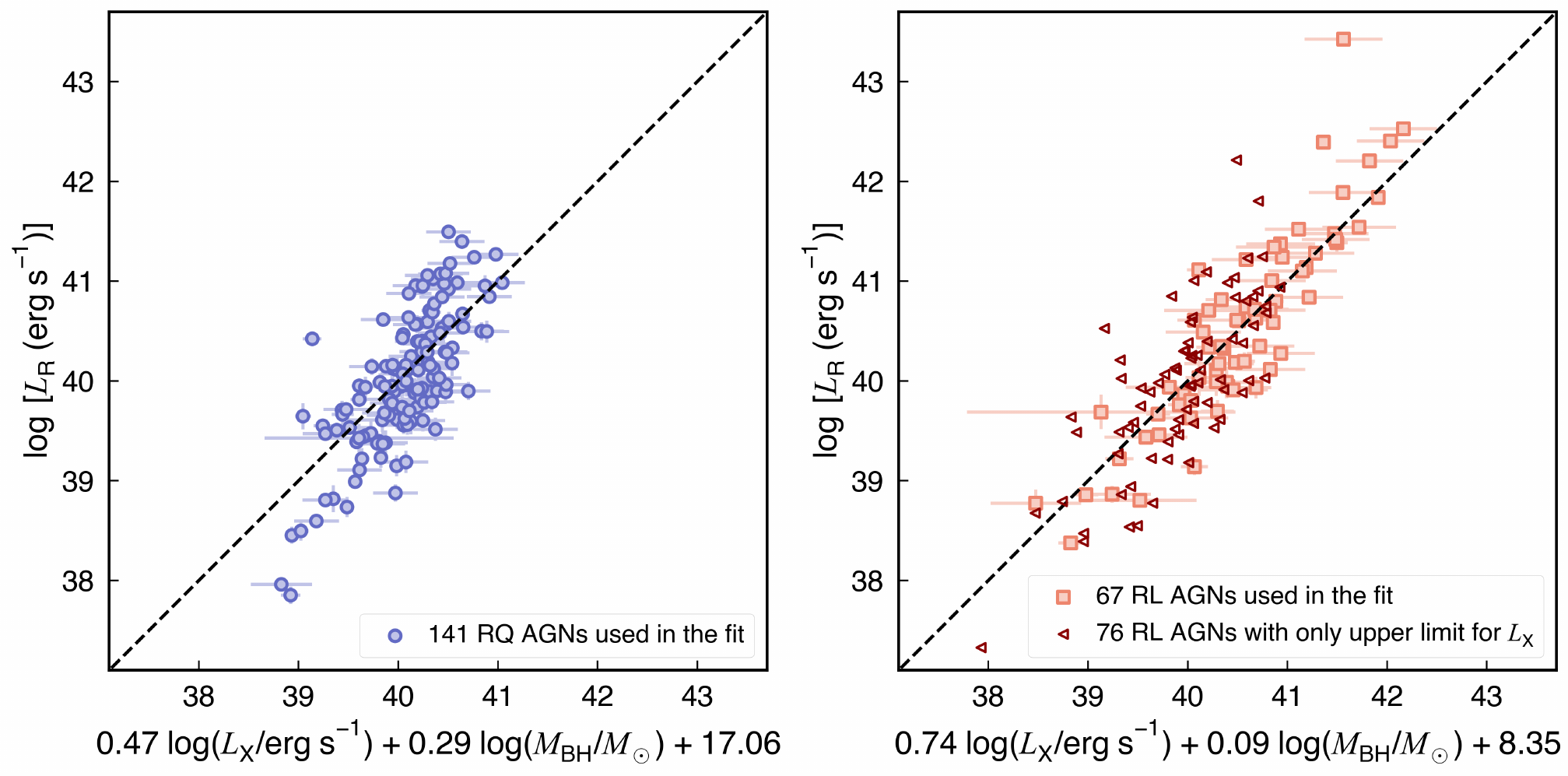}
\caption{Comparison of the predicted rest-frame 5 GHz luminosity from the best-fit fundamental plane ({\it x axis}) and the observed rest-frame 5 GHz luminosity ({\it y axis}) for the radio-quiet AGNs ({\it left panel}) and the radio-loud AGNs ({\it right panel}). The dashed black line in each panel represents the 1:1 line. The empty dark red triangles in the right panel represent the radio-loud AGNs with only upper limits for $L_{\rm X}$ (see parameter calculation in Section \ref{sec:LXpara} and see sample selection in Section \ref{sec:fitness}) that were not utilized in the regression fitting for the fundamental plane.
\label{fig:BHFPfit}}
\end{figure*}

\begin{itemize}
    \item[(1)] Both the radio-loud and radio-quiet AGN subsamples have a similar redshift range between 0.1 and 4, while the median redshift is 0.9 for the radio-loud AGNs and 1.2 for the radio-quiet AGNs (see Panel A of Fig. \ref{fig:RLRQproperties}). The $p$ value of the K-S test for the redshift distributions of these two subsamples is about 0.037, which is lower than 0.05. This means that the radio-loud and radio-quiet AGN subsamples have different redshift distributions.
    \item[(2)] The range of $\log [L_{\rm X}\ (\rm{erg\ s}^{-1})$] for our radio-quiet AGN sample is between $40.9$ and $45.2$ with a median of $44.0$, while for our radio-loud AGN sample it is between $39.9$ and $44.7$ with a median of $42.4$ (see Panel B of Fig. \ref{fig:RLRQproperties}). The radio-quiet AGN sample has significantly larger $L_{\rm X}$ than the radio-loud AGN sample. The $p$ value of the K-S test is about $10^{-16}$; that is, much lower than 0.05, which indicates a significantly different X-ray luminosity distribution between the radio-loud and radio-quiet AGN subsamples.
    \item[(3)] The range of $\log [L_{\rm R}\ (\rm{erg\ s}^{-1})$] for the radio-quiet AGN sample is between $37.9$ and $41.5$ with a median of $40.0$, while for the radio-loud AGN sample it is between $38.4$ and $43.4$ with a median of $40.5$ (see Panel C of Fig. \ref{fig:RLRQproperties}). The radio-quiet AGN sample has slightly lower $L_{\rm R}$ than the radio-loud AGN sample. The $p$ value of the K-S test is about $0.0006$; that is, much lower than 0.05, which indicates a significantly different radio luminosity distribution between the radio-loud and radio-quiet AGN subsamples.
    \item[(4)] The range of the specific star formation rate (sSFR) for our radio-quiet AGN sample is between $0.001$ and $100$ Gyr$^{-1}$ with a median of 0.60 Gyr$^{-1}$, while for our radio-loud AGN sample it is between $0.003$ and 17 Gyr$^{-1}$ with a median of 0.12 Gyr$^{-1}$ (see Panel D of Fig. \ref{fig:RLRQproperties}). Here, the SFR was derived from Paper I, which estimated it with the broadband SED fitting. The radio-quiet AGN sample has a slightly higher sSFR than the radio-loud AGN sample. The $p$ value of the K-S test is about $0.0008$; that is, much lower than 0.05, which demonstrates that the radio-loud and radio-quiet AGN subsamples have significantly different sSFRs. This conclusion can also be proved by the different SFG or QG proportions in these two subsamples (see Figure \ref{fig:LRLXdis} or Section \ref{sec:FPgal}).
    \item[(5)] The range of $\log [M_{\rm BH}/M_{\odot}]$ for the radio-quiet AGN sample is between $5.35$ and $9.62$ with a median of $8.36$, while for the radio-loud AGN sample it is between $4.28$ and $9.70$ with a median of $8.58$ (see Panel E of Fig. \ref{fig:RLRQproperties}). The $p$ value of the K-S test is about $0.122$; that is, higher than 0.05, which indicates that the radio-loud and radio-quiet AGN subsamples have a similar black hole mass.
    \item[(6)] The Eddington ratio ($\lambda_{\rm Edd}$) was defined as $\lambda_{\rm Edd}=L_{\rm bol}/L_{\rm Edd}$, where $L_{\rm bol}$ is the bolometric luminosity and $L_{\rm Edd}$ is the Eddington luminosity.
The bolometric luminosity is converted from X-ray luminosity by a bolometric correction factor that is related to the luminosity \citep[e.g.,][]{Duras2020}.
For our sample, this correction factor was around 20 \citep[][]{Duras2020}. Therefore, here $L_{\rm bol}$ was calculated by $L_{\rm bol} = L_{\rm X}\times 20$.
The Eddington luminosity was defined as $L_{\rm Edd} = 1.3 \times 10^{38}\ M_{\rm BH}/M_{\odot}$ in units of erg s$^{-1}$ \citep{Rybicki1979}.
The range of $\lambda_{\rm Edd}$ for our radio-quiet AGN sample is between $2.3\times 10^{-5}$ and $51$ with a median of $0.06$, while for our radio-loud AGN sample it is between $10^{-5}$ and $1.0$ with a median of $0.001$ (see Panel F of Fig. \ref{fig:RLRQproperties}). The radio-quiet AGN sample has significantly larger $\lambda_{\rm Edd}$ than the radio-loud AGN sample, which is consistent with the negative correlation between radio loudness and the Eddington ratio \citep[e.g.,][]{Ho2002,Panessa2007,Sikora2007}. The $p$ value of the K-S test is about $10^{-12}$; that is, significantly lower than 0.05, which indicates that the radio-loud and radio-quiet AGN subsamples present significantly different Eddington ratios.
\end{itemize}

\begin{table*}[!t]
 \centering
 \setlength{\tabcolsep}{10pt}
 \caption{Best-fitting parameters of the fundamental plane of black hole activity for radio-quiet and radio-loud AGNs in this work. \label{tab:BHFPparameter}}
 \begin{tabular}{cccccccccccc}
 \hline\hline\xrowht[()]{10pt}
 & $N_{\rm sources}$ & $\xi_{\rm X}$ & $\xi_{\rm M}$ & $b$ & $\sigma_{\rm R}$ \\
 \hline\xrowht[()]{10pt}
 & & \multicolumn{2}{c}{Radio-quiet AGNs} & & \\
  \hline\xrowht[()]{5pt}
Total & 141 & $0.47 \pm 0.06$ & $0.29 \pm 0.06$ & $17.06^{+2.45}_{-2.47}$ & 0.39  \\
 & & & & & \\
 \xrowht[()]{5pt}
$\lambda_{\rm Edd} \leq 10^{-1.5}$ & 63 & $0.49 \pm 0.10$ & $0.44 \pm 0.17$ & $14.76 \pm 3.75$ & 0.33  \\
\xrowht[()]{5pt}
$\lambda_{\rm Edd} > 10^{-1.5}$ & 78 & $0.35 \pm 0.13$ & $0.30 \pm 0.09$ & $22.14 \pm 5.42$ & 0.43  \\
 & & & & & \\
 \xrowht[()]{5pt}
$0.1< z \leq 1.0$ & 56 & $0.45 \pm 0.08$ & $0.26 \pm 0.09$ & $17.78^{+3.65}_{-3.66}$ &  0.37 \\
\xrowht[()]{5pt}
$1.0< z \leq 4.0$ & 85 & $0.28 \pm 0.09$ & $0.31 \pm 0.08$ & $25.57^{+3.84}_{-3.81}$ &  0.38 \\
 & & & & & \\
 \xrowht[()]{5pt}
Hosted by SFGs & 116 & $0.36 \pm 0.07$ & $0.32 \pm 0.07$ & $21.79^{+3.20}_{-3.22}$ &  0.41 \\
\xrowht[()]{5pt}
Hosted by QGs & 25 & $0.63 \pm 0.08$ & $0.18 \pm 0.10$ & $11.05^{+3.63}_{-3.64}$ & 0.23  \\
\hline\xrowht[()]{10pt}
& & \multicolumn{2}{c}{Radio-loud AGNs} & & \\
  \hline\xrowht[()]{5pt}
Total & 67 & $0.74 \pm 0.09$ & $0.09 \pm 0.09$ & $8.35^{+3.61}_{-3.60}$ & 0.37  \\
 & & & & & \\
 \xrowht[()]{5pt}
$\lambda_{\rm Edd} \leq 10^{-2.5}$ & 40 & $0.44 \pm 0.14$ & $0.34 \pm 0.19$ & $18.60^{+5.27}_{-5.20}$ &  0.36 \\
\xrowht[()]{5pt}
$\lambda_{\rm Edd} > 10^{-2.5}$ & 27 & $0.72 \pm 0.21$ & $0.20 \pm 0.19$ & $8.67^{+8.10}_{-8.08}$ & 0.40  \\
 & & & & & \\
 \xrowht[()]{5pt}
$0.1< z \leq 1.0$ & 37 & $0.71 \pm 0.11$ & $0.06 \pm 0.11$ & $9.77^{+4.45}_{-4.47}$ &  0.34 \\
\xrowht[()]{5pt}
$1.0< z \leq 4.0$ & 30 & $0.74 \pm 0.17$ & $0.23 \pm 0.21$ & $7.21^{+7.48}_{-7.07}$ &  0.39 \\
 & & & & & \\
 \xrowht[()]{5pt}
Hosted by SFGs & 32 & $0.72 \pm 0.14$ & $0.17 \pm 0.14$ & $8.44^{+5.50}_{-5.43}$ &  0.37 \\
\xrowht[()]{5pt}
Hosted by QGs & 35 & $0.73 \pm 0.11$ & $-0.07 \pm 0.20$ & $9.84^{+4.88}_{-4.92}$ & 0.35  \\
 \hline
 \end{tabular}
 \tablefoot{
 $N_{\rm sources}$ represents the number of sources. The best-fit parameters for the fundamental plane correspond to the 
 correlation coefficient of X-ray luminosity, $\xi_{\rm X}$, the correlation coefficient of black hole mass, $\xi_{\rm M}$, and constant offset, $b$ (see Equation \ref{eq:BHFP}).
 The parameter $\sigma_{\rm R}$ denotes the scatter of the observed $L_{\rm R}$ relative to the predicted value from the best-fit fundamental plane.
 \label{tab:fitness}
 }
 \end{table*}

\section{Fitting the fundamental plane for radio-loud and radio-quiet active galactic nuclei}
\label{sec:fitness}

Following \cite{Merloni2003}, we defined the fundamental plane of black hole activity as
\begin{equation}
\log L_{\rm R} = \xi_{\rm X}\log L_{\rm X}+\xi_{\rm M}\log M_{\rm BH}+b,
\label{eq:BHFP}
\end{equation}
where $L_{\rm R}$ is the rest-frame 5 GHz radio luminosity (in units of erg s$^{-1}$),
$L_{\rm X}$ is the rest-frame 2--10 keV X-ray luminosity (in units of erg s$^{-1}$),
and $M_{\rm BH}$ is the black hole mass (in units of $M_{\odot}$).
Then we used the the ordinary least squares (OLS) linear regression method in the Python package \textsc{scikit-learn} \citep{Pedregosa2011}
to fit the dataset. 
Further, we ran the fitting process with 4000 iterations to take the uncertainties of both dependent and independent variables into account.
Next, we took the variable $L_{\rm X}$ as an example to show the detailed analysis process.
In our sample, the $i$th object has the observed $L_{{\rm X,obs},i}$ with an uncertainty of $\sigma_i$.
In the $j$th fitting process, the value of the variable used for fitting ($L_{{\rm X,fit},i,j}$) was randomly selected from a normal distribution 
centered at $L_{{\rm X,obs},i}$ with a variance of $\sigma_i$.
We made the same value selection for the variables $L_{\rm R}$ and $M_{\rm BH}$.
Thus, in the $j$th fitting process, the $i$th object had $L_{{\rm X,fit},i,j}$, $L_{{\rm R,fit},i,j}$, and $M_{{\rm BH,fit},i,j}$.
After the $j$th fitting, the best-fit parameters were given as $\xi_{{\rm X,} j}\pm \sigma_{{\rm X,} j}$, $\xi_{{\rm M,} j}\pm \sigma_{{\rm M,} j}$, and $b_{\rm j}\pm \sigma_{b,j}$. Finally, fitting with 4000 iterations produced a nearly symmetric distribution for each fitting coefficient.
Then we used the mean of the distribution as the best-fit coefficient, while the error was derived from
the integrated variance from both fitting process and observed uncertainties. Next, we took the parameter $\xi_{\rm X}$ as an example.
The best-fit $\xi_{\rm X}$ was calculated by $\sum_{j=1}^{4000} \xi_{{\rm X,} j}/4000$ (written as $\bar{\xi}_{\rm X}$ hereafter), while its error was calculated by $\sqrt{\sum_{j=1}^{4000} (\xi_{{\rm X,} j}-\bar{\xi}_{\rm X})^2/4000+\left(\sum_{j=1}^{4000} \sigma_{{\rm X,} j}/4000\right)^2}$. Then we did the same calculation for both $\xi_{\rm M}$ and $b$.

We performed the above fitting process for our radio-quiet and radio-loud AGN samples separately.
We found a best-fit fundamental plane of  
\begin{equation}
\log L_{\rm R} = (0.47\pm 0.06)\log L_{\rm X}+(0.29\pm 0.06)\log M_{\rm BH}+17.06^{+2.45}_{-2.47}
\end{equation}
for the radio-quiet AGN sample, and
\begin{equation}
\log L_{\rm R} = (0.74\pm 0.09)\log L_{\rm X}+(0.09\pm 0.09)\log M_{\rm BH}+8.35^{+3.61}_{-3.60}
\label{eq:RLFP}
\end{equation}
for the radio-loud AGN sample.
The best-fit coefficients and their errors are summarized in Table \ref{tab:BHFPparameter}.
The comparison between the observed $L_{\rm R}$ and the predicted $L_{\rm R}$ by the best-fit fundamental plane is shown in Fig. \ref{fig:BHFPfit}.
The significantly different fundamental plane between radio-quiet and radio-loud AGNs, and their comparison with literature works, are discussed in detail in Section \ref{sec:FPstudies}.
Here, we defined a parameter, $\sigma_{\rm R}$, to denote the scatter of the observed $L_{\rm R}$ relative to the predicted value from the best-fit fundamental plane. $\sigma_{\rm R}$ for the radio-loud and radio-quiet subsets was 0.37 and 0.39, respectively (see Table \ref{tab:BHFPparameter}). We also fit the fundamental plane for the entire radio AGN sample, which shows a much larger scatter with $\sigma_{\rm R}=0.64$.
Next, we calculated the $\chi^2$ and degree of freedom ($dof$) for the best-fit results of the entire sample and the two subsets, respectively. 
Then we used $f$-test at a confidence level of $\geq$ 95\% to decide whether splitting the entire sample into the radio-loud and radio-quiet subsets gives different coefficients for the fundamental plane. The $f$-test $p$ values based on the $\chi^2$ and $dof$ of the entire sample and each subset are all much lower than 0.05. Therefore, these statistical results verify that the radio-loud and radio-quiet AGNs in our sample do follow different fundamental planes.
For the radio AGNs with only upper limits for X-ray luminosity, we estimated the lower limits of their radio loudness according to the ratio of $L_{\rm R}$ to $L_{\rm X}$ upper-limits. Therefore, from these sources, we selected 76 radio-loud AGNs with only upper limits for X-ray luminosity based on the previously defined radio-loudness threshold (see Equation \ref{eq:radioloudness}). These objects were not included in the regression fitting for the fundamental plane, and they are just shown in Fig. \ref{fig:BHFPfit} (see the dark red triangles) as a comparison. These radio-loud AGNs are greatly consistent with the fundamental plane (see Equation \ref{eq:RLFP}) obtained with the aforementioned radio-loud AGN sample.
 
\begin{figure*}
\includegraphics[width=\linewidth, clip]{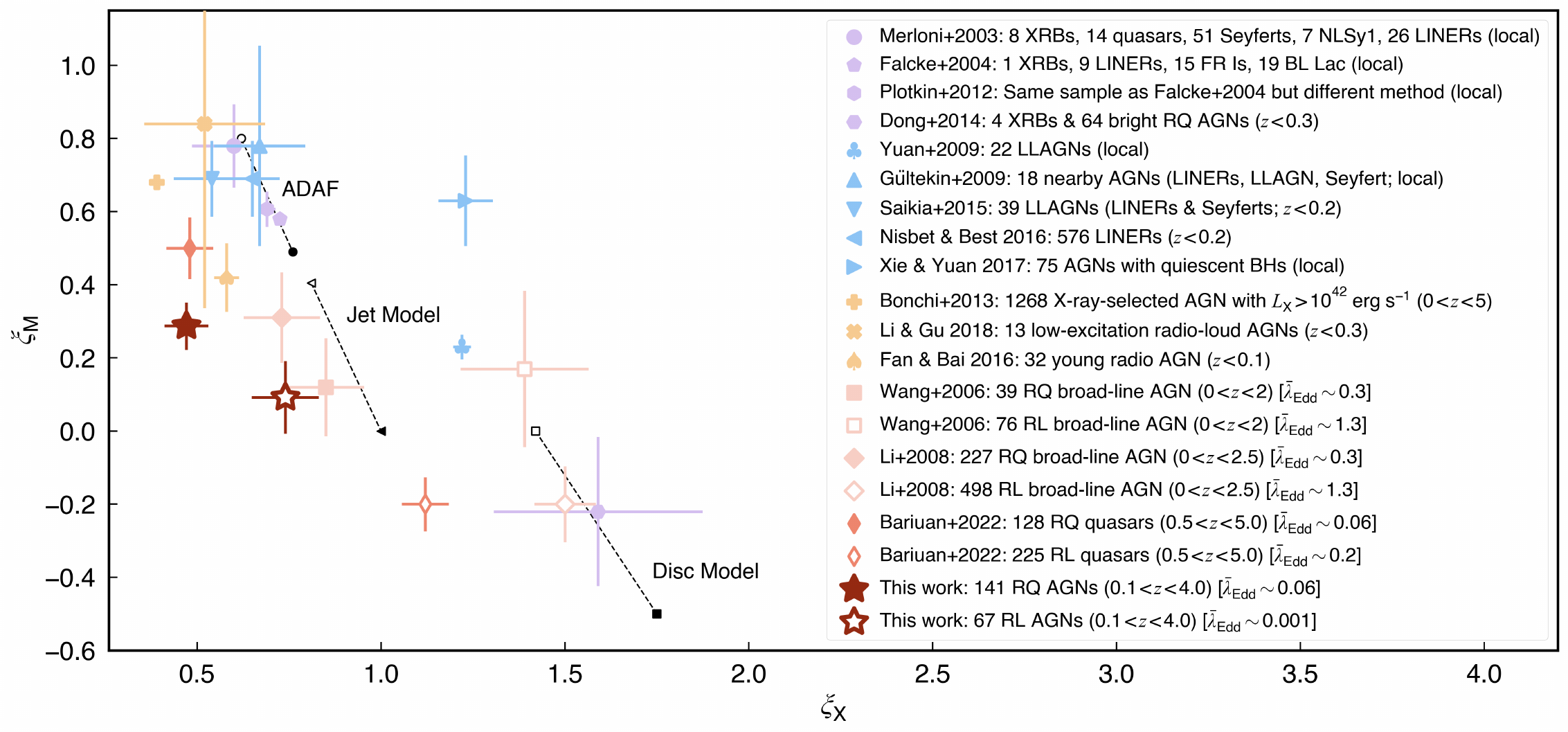}
\caption{Comparison of the best-fit correlation coefficients, $\xi_{X}$ and $\xi_{M}$, of the fundamental plane. Our results for the radio-quiet (RQ) AGNs and radio-loud (RL) AGNs samples are shown as the filled dark red star and the empty dark red star, respectively. The results from literature works are shown as comparisons, including \cite{Merloni2003} (Merloni+2003), \cite{Falcke2004} (Falcke+2004), \cite{Plotkin2012} (Plotkin+2012), \cite{Dong2014} (Dong+2014), \cite{Yuan2009} (Yuan+2009), \cite{Gultekin2009} (G$\rm \ddot{u}$ltekin+2009), \cite{Saikia2015} (Saikia+2015), \cite{Nisbet2016} (Nisbet \& Best 2016), \cite{Xie2017} (Xie \& Yuan 2017), \cite{Bonchi2013} (Bonchi+2013), \cite{Li2018} (Li \& Gu 2018), \cite{Fan2016} (Fan \& Bai 2016), \cite{Wang2006} (Wang+2006), \cite{Li2008} (Li+2008), and \cite{Bariuan2022} (Bariuan+2022). 
Purple symbols represent studies based on a sample consisting of both XRBs and local AGNs. Blue symbols represent studies based on LLAGNs. Yellow symbols represent studies for other types of AGNs. Red symbols represent studies for radio-quiet (filled symbols) and radio-loud (empty symbols) AGNs separately.
The theoretically predicted correlation coefficients were derived from \cite{Merloni2003} with the electron spectral index $p=2$, and are shown as black circles for the ADAF, black triangles for a synchrotron jet model, and black squares for the standard Shakura-Sunyaev disc model. Among them, empty and filled symbols represent the predictions based on a radio spectral index of $0$ and $-0.5$, respectively, and the dashed lines connecting the empty and filled symbols represent the tracks of $\xi_{\rm X}$ and $\xi_{\rm M}$ due to the variation in the radio spectral index. 
\label{fig:RXRMCCslope}}
\end{figure*}

\section{Discussion}
\label{sec:discussion}

The comparison between the best-fit correlation coefficients, $\xi_{\rm X}$ and $\xi_{\rm M}$, of the fundamental plane can characterize the accretion physics and origins of X-ray emission for central engines \citep{Merloni2003}.
In the $\xi_{\rm X}$--$\xi_{\rm M}$ diagram (see Fig. \ref{fig:RXRMCCslope}, Fig. \ref{fig:RXRMCCproperties}, and Fig. \ref{fig:cartoon}), the theoretically predicted correlation coefficients were derived from \cite{Merloni2003}. These theoretical models are advection-dominated accretion flow \citep[ADAF; a radiatively inefficient accretion flow;][for a review]{Narayan1994,Yuan2001,Yuan2014}, the synchrotron jet model producing radio emission and X-ray emission by the optically thin synchrotron radiation from jet, and the standard Shakura-Sunyaev disc model \citep[][for a review]{Shakura1973,Pringle1981}. 
We followed \cite{Merloni2003} in assuming an electron spectral index of $p=2$ for all the models, and assumed an accretion efficiency coefficient of $q=2.3$ for the ADAF model and $q=1$ for the disc model.
Next, we give a brief summary of the studies on the fundamental plane in Section \ref{sec:FPstudies} and further discuss the dependence of the fundamental plane on Eddington ratio (Section \ref{sec:FPEdd}), redshift (Section \ref{sec:FPz}), and galaxy star formation properties (Section \ref{sec:FPgal}). Combining our and literature works, we investigate the central engines of radio-quiet and radio-loud AGNs in Section \ref{sec:engine}.

\subsection{A brief summary of the fundamental plane studies}
\label{sec:FPstudies}

\cite{Merloni2003}, \cite{Falcke2004}, \cite{Plotkin2012}, and \cite{Dong2014} all studied the fundamental plane using a sample consisting of both XRBs and local AGNs (see the purple symbols in Fig. \ref{fig:RXRMCCslope}). 
The AGN samples of \cite{Merloni2003}, \cite{Falcke2004}, and \cite{Plotkin2012} mainly focused on low-luminosity AGNs (LLAGNs).
\cite{Merloni2003} favor the X-ray emission being produced by an ADAF model, while \cite{Falcke2004} and \cite{Plotkin2012} support the X-ray emission being produced by synchrotron radiation from the inner jet based on a sample having flat or inverted spectra.
On the other hand, \cite{Dong2014} showed a significantly different pattern, that their bright AGNs with higher Eddington ratio are in line with the predictions of the disc model. Such a difference indicates that different types of AGNs may have different fundamental planes.

Furthermore, many studies of the fundamental plane only focus on AGNs. 
Some works studied the fundamental plane of local LLAGNs \citep[][see blue symbols in Fig. \ref{fig:RXRMCCslope}]{Yuan2009,Gultekin2009b,Saikia2015,Nisbet2016,Xie2017}. \cite{Gultekin2009b}, \cite{Saikia2015}, and \cite{Nisbet2016} obtained consistent results with those in \cite{Merloni2003} and \cite{Falcke2004}, while \cite{Yuan2009} and \cite{Xie2017} exhibited different results. 
The LLAGN samples in \cite{Gultekin2009b}, \cite{Saikia2015}, and \cite{Nisbet2016} have a higher Eddington ratio than those in \cite{Yuan2009} and \cite{Xie2017}, which might explain the different results between them. 
X-ray-selected AGNs ($L_{\rm X} > 10^{42}\ \rm{erg}\ \rm{s}^{-1}$) from \cite{Bonchi2013} and young radio AGNs from \cite{Fan2016} show a slight deviation from the ADAF model.
\cite{Bonchi2013} mentioned that the deviation is mainly due to the inclusion of the radio upper limits in their analysis,
while \cite{Fan2016} thought that the bad angular resolution of radio data cannot distinguish the extended lobes from the cores, which will result in flatter correlation coefficients.
The low-excitation radio-loud AGNs from \cite{Li2018} show a similar fundamental plane with the LLAGNs (see Fig. \ref{fig:RXRMCCslope}).
The above works usually calculated luminosity assuming a different radio spectral index, which may also bias the fundamental plane.

In addition, many works explore the fundamental plane of radio-quiet and radio-loud populations separately \citep{Wang2006,Li2008,Bariuan2022}.
Both \cite{Wang2006} and \cite{Li2008} utilized broad-line AGN samples and they got consistent results for the fundamental plane within a 1$\sigma$ uncertainty (see Fig. \ref{fig:RXRMCCslope}). They all found that radio-quiet and radio-loud AGNs show quite different fundamental planes. Radio-quiet AGNs favor the ADAF model coupled with a synchrotron jet model, while the radio-loud AGNs favor a combination of the disc model with the synchrotron jet model.
More recently, \cite{Bariuan2022} found that for luminous quasars, radio-quiet and radio-loud subsets also have different fundamental planes. Therefore, these works demonstrate that the fundamental plane depends on the radio loudness of AGNs.
The rest-frame 5 GHz luminosities of \cite{Wang2006}, \cite{Li2008}, and \cite{Bariuan2022} were all converted from 1.4 GHz radio flux densities assuming a radio spectral index of $-0.5$.

We also separately studied the fundamental plane of radio-quiet and radio-loud AGNs. Compared to the literature works \citep{Wang2006,Li2008,Bariuan2022}, our samples have a lower X-ray luminosity, lower radio luminosity, and lower Eddington ratio. We also find that there is a different fundamental plane between radio-quiet and radio-loud AGNs. Our low-luminosity radio-quiet AGNs show a slightly different fundamental plane than the high-luminosity radio-quiet AGNs from \cite{Wang2006}, \cite{Li2008}, and \cite{Bariuan2022}, while our low-luminosity radio-loud AGNs present a significantly different result than the high-luminosity radio-loud AGNs from these literature works (see Fig. \ref{fig:RXRMCCslope}). These results indicate that the fundamental plane may have a dependence on the Eddington ratio (see details in Section \ref{sec:FPEdd}).

\begin{figure*}[th!]
\includegraphics[width=\linewidth, clip]{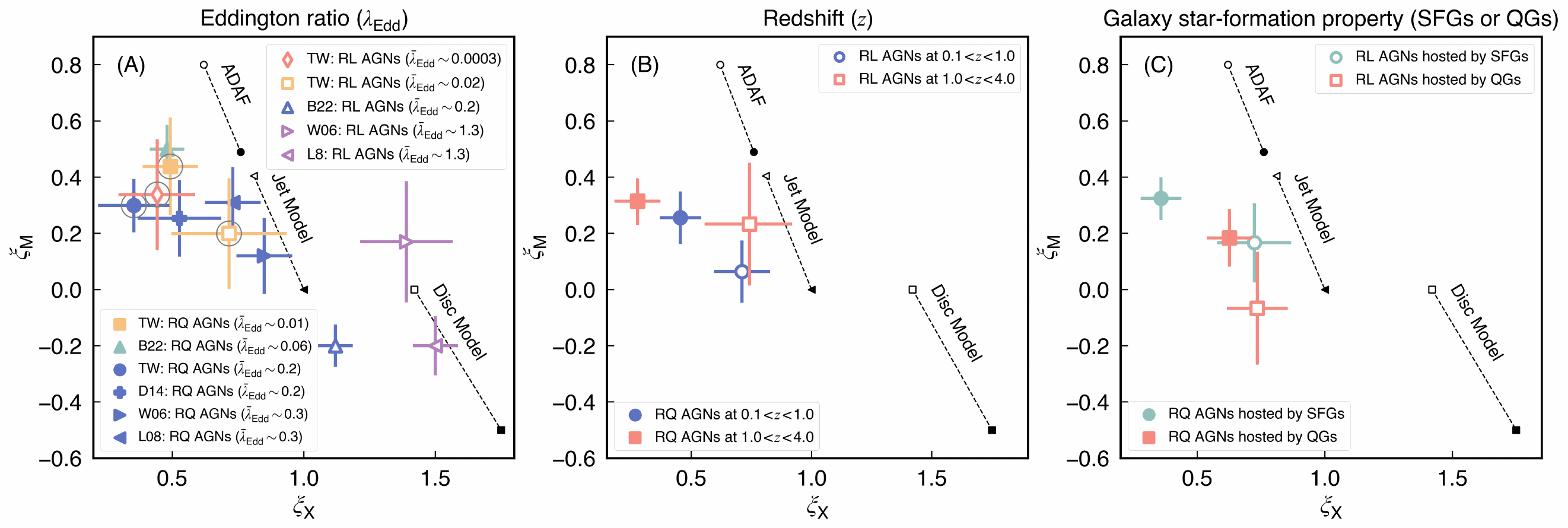}
\caption{Comparison of the best-fit correlation coefficients, $\xi_{X}$ and $\xi_{M}$, of the fundamental plane for Eddington ratio subsamples (A), redshift subsamples (B), and galaxy star formation property subsamples (C). The theoretically predicted correlation coefficients for ADAF, the synchrotron jet model, and the disc model are the same as in Fig. \ref{fig:RXRMCCslope}, and are shown here as circles, triangles, and squares, respectively. 
In each panel, filled symbols represent radio-quiet (RQ) AGNs and empty symbols represent radio-loud (RL) AGNs.
In Panel (A), ``TW'', ``B22'', ``D14'', ``W06'', and ``L08'' mean results from this work, \cite{Bariuan2022}, \cite{Dong2014}, \cite{Wang2006}, and \cite{Li2008}, respectively.
Different colors represent different Eddington ratios: red for $\bar{\lambda}_{\rm Edd} < 0.01$, yellow for $0.01 < \bar{\lambda}_{\rm Edd} < 0.05$, green for $0.05 < \bar{\lambda}_{\rm Edd} < 0.1$, blue for $0.1 < \bar{\lambda}_{\rm Edd} < 0.5$, and violet for $\bar{\lambda}_{\rm Edd} > 0.5$. In addition, our results (TW) are labeled by big gray circles.
\label{fig:RXRMCCproperties}}
\end{figure*}

The best-fit $\xi_{\rm M}$ coefficients of \cite{Wang2006}, \cite{Li2008}, \cite{Fan2016}, \cite{Bariuan2022}, and our results appear relatively small, and fall below the theoretical predications in Fig. \ref{fig:RXRMCCslope}, especially for the radio-quiet subsamples at a low Eddington ratio (see Panel A of Figure \ref{fig:RXRMCCproperties}). 
The theoretical models shown in Fig. \ref{fig:RXRMCCslope} were obtained based on the assumption of $p=2$. We calculated the theoretical models with $p=3$, but they still cannot explain the observations.
Thus, we performed two fitting tests aiming to discuss possible reasons. Firstly,
we simultaneously fit the XRBs and LLAGNs from \cite{Merloni2003}, our sample, and the quasar sample from \cite{Bariuan2022}. The luminosity range of our sample is between the LLAGNs of \cite{Merloni2003} and the quasar sample of \cite{Bariuan2022}. The best-fit $\xi_{\rm M}$ and $\xi_{\rm X}$ are $0.59 \pm 0.04$ and $0.85 \pm 0.03$, respectively ($\sigma_{\rm R} \sim 0.7$). Next, we only fit the LLAGNs from \cite{Merloni2003}, our sample, and the quasar sample from \cite{Bariuan2022}. The best-fit $\xi_{\rm M}$ and $\xi_{\rm X}$ are $0.57 \pm 0.07$ and $0.86 \pm 0.03$, respectively ($\sigma_{\rm R} \sim 0.7$). 
Both of these two tests give results consistent with \cite{Merloni2003}, which implies that the X-ray emission is produced by the ADAF model. 
These results differ from those using our sample and the quasar sample \citep{Bariuan2022}, respectively. 
From a statistical perspective, linear regressions to noisy data with limited dynamic range may result in different coefficients than that with a broad dynamic range \citep{Plotkin2012}.
However, given that we use a large sample with a broad luminosity range from LLAGNs to quasars, we do not expect that they follow similar accretion physics and X-ray radiation mechanisms. 
Therefore, from a scientific perspective, the observational deviation from the theoretical predictions may indicate that some other underlying physical properties have an effect on the fundamental plane. 

In addition, other mechanisms may be also responsible for the observational deviation from the theoretical predictions, especially at low Eddington ratios; such as:
1) We only use one solution (i.e., ADAF) of radiatively inefficient flows following \cite{Merloni2003}, but we cannot rule out the possibility that other solutions, such as convection-dominated accretion flows, will better explain the observations;
2) Radiatively inefficient flows are prone to produce powerful outflows that may alter the theoretical predictions significantly \citep{Merloni2003};
3) For the most massive SMBHs ($>\ \sim 10^8\ M_{\odot}$), the synchrotron cooling in jet becomes important, which may bias the fundamental plane \citep{Falcke2004, Kording2006, Plotkin2012}.
Even so, it is still meaningful to investigate whether other physical properties or processes, such as the Eddington ratio (accretion rate), redshift (cosmic evolution), and galaxy star formation properties (fueling mechanisms) affect the fundamental plane.

\subsection{Eddington-ratio dependence of the fundamental plane}
\label{sec:FPEdd}
We split our radio-quiet AGN subsample into two Eddington-ratio subsets: $\lambda_{\rm Edd} \leq 10^{-1.5}$ (with a median value of $0.01$) and $\lambda_{\rm Edd} > 10^{-1.5}$ (with a median value of $0.2$). Our radio-loud AGN subsample was split into two Eddington-ratio subsets: $\lambda_{\rm Edd} \leq 10^{-2.5}$ (with a median value of $0.0003$) and $\lambda_{\rm Edd} > 10^{-2.5}$ (with a median value of $0.02$).
The best-fit parameters for the fundamental plane of these subsets and the source number in each subset are summarized in Table \ref{tab:BHFPparameter}. The comparison between $\xi_{\rm X}$ and $\xi_{\rm M}$ is shown in Panel A of Fig. \ref{fig:RXRMCCproperties}.
For the radio-quiet AGN sample, both the low-Eddington-ratio subset ($\lambda_{\rm Edd} \leq 10^{-1.5}$; the filled yellow square in Panel A of Fig. \ref{fig:RXRMCCproperties}) and high-Eddington-ratio subset ($\lambda_{\rm Edd} > 10^{-1.5}$; the filled blue circle in Panel A of Fig. \ref{fig:RXRMCCproperties}) agree with an ADAF model coupled with a synchrotron jet model for the origins of X-ray emission. For the radio-loud AGN sample, X-ray emission origins of the low-Eddington-ratio subset ($\lambda_{\rm Edd} \leq 10^{-2.5}$; the empty red diamond in Panel A of Fig. \ref{fig:RXRMCCproperties}) can be explained by an ADAF model coupled with a synchrotron jet model, while the high-Eddington-ratio subset ($\lambda_{\rm Edd} > 10^{-2.5}$; the empty yellow square in Panel A of Fig. \ref{fig:RXRMCCproperties}) mainly follows a synchrotron jet model.
For comparison, $\xi_{\rm X}$ and $\xi_{\rm M}$ from \cite{Wang2006} (W06), \cite{Li2008}  (L08), and \cite{Bariuan2022}  (B22) are shown in Panel A of Fig. \ref{fig:RXRMCCproperties}.
In addition, we also analyzed the fundamental plane for the radio-quiet AGN sample from \cite{Dong2014} (D14) whose best-fit parameters are also shown in Panel A of Fig. \ref{fig:RXRMCCproperties}.
Our samples and samples from these literature works have different Eddington ratios and follow different fundamental planes (see Panel A of Fig. \ref{fig:RXRMCCproperties}).
According to these results, we find that for both radio-quiet AGNs and radio-loud AGNs, their fundamental planes may depend on Eddington ratio.
We stress that only for our radio-loud and radio-quiet AGN samples does the scatter, $\sigma_{\rm R}$, not show significant improvement after splitting the sample into two Eddington-ratio subsets (see Table \ref{tab:BHFPparameter}), and the $f$-test $p$ values based on the $\chi^2$ and $dof$ of the entire sample and each Eddington-ratio subset are higher than 0.05. Therefore, only for our samples do the statistical results imply no significant dependence of the fundamental plane on the Eddington ratio, while combining with the quasar samples from previous works, this dependence becomes significant.
It demonstrates that using a sample including both low-luminosity and luminous AGNs with a broad Eddington-ratio range is necessary for us to draw a solid conclusion about the dependence of the fundamental plane on the Eddington ratio.

\begin{figure*}
\centering
\includegraphics[width=0.9\linewidth, clip]{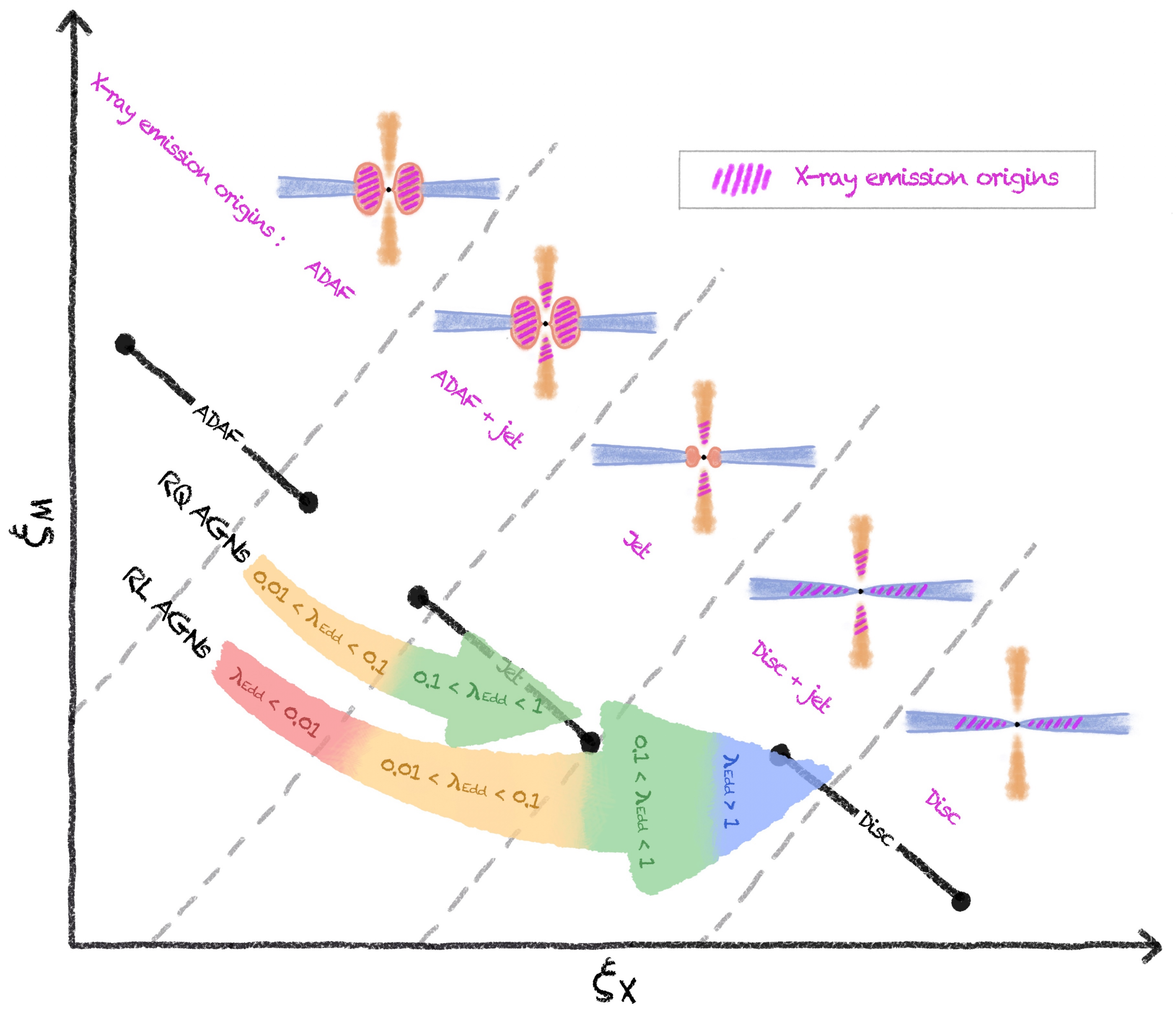}
\caption{Cartoon illustrating the model for the central engine of radio-quiet AGNs and radio-loud AGNs. Solid black lines represent the theoretically predicted correlation coefficients for ADAF, jet, and standard thin disc models, which are same as Fig. \ref{fig:RXRMCCslope} and Fig. \ref{fig:RXRMCCproperties}.
The upper colored arrow represents the observational results for radio-quiet (RQ) AGNs, 
while the lower colored arrow represents the observational results for radio-loud (RL) AGNs.
The different colors of the arrow represent different Eddington ratios ($\lambda_{\rm Edd}$)
and the direction of the arrow means from low $\lambda_{\rm Edd}$ to high $\lambda_{\rm Edd}$.
The shaded violet regions represent the origins of X-ray emission.
The origins of X-ray emission of radio-quiet AGNs at $0.01 < \lambda_{\rm Edd} < 0.1$ are consistent with a combination of ADAF and a synchrotron jet model (ADAF+jet model), while at $0.1 < \lambda_{\rm Edd} < 1$, they mainly follow the synchrotron jet model (Jet model). 
The origins of X-ray emission of radio-loud AGNs are consistent with a combination of ADAF and a synchrotron jet model (ADAF+jet model) at $\lambda_{\rm Edd} < 0.01$, and agree with the synchrotron jet model (Jet model) at $0.01 < \lambda_{\rm Edd} < 0.1$, and follow the standard thin disc coupled with a synchrotron jet model (Disc+jet model) at $\lambda_{\rm Edd} > 0.1$.
\label{fig:cartoon}}
\end{figure*}

\subsection{Redshift dependence of the fundamental plane}
\label{sec:FPz}
We further studied the fundamental plane of radio-quiet and radio-loud AGN subsamples by splitting them into two redshift subsets: $0.1 < z \leq 1$ and $1 < z \leq 4$. For each subset, the best-fit parameters and source numbers are summarized in Table \ref{tab:BHFPparameter}. 
Splitting the radio-loud or radio-quiet AGN sample into two redshift subsets does not reduce the scatter, $\sigma_{\rm R}$ (see Table \ref{tab:BHFPparameter}), and produces $f$-test $p$ values based on the $\chi^2$ and $dof$ higher than 0.05. Therefore, neither the radio-loud AGNs nor the radio-quiet AGNs show a statistically significant redshift dependence of the fundamental plane (also see Panel B of Fig. \ref{fig:RXRMCCproperties}).
For both radio-loud AGNs and radio-quiet AGNs, there seems to be no significant redshift dependence of the fundamental plane (see Panel B of Fig. \ref{fig:RXRMCCproperties}).
For luminous AGNs (such as broad-line AGNs and quasars), \cite{Li2008} found no significant dependence on redshift for both radio-quiet and radio-loud subsets across $0 < z < 2.5$, while \cite{Bariuan2022} only observed a redshift dependence for radio-loud subsets across $0.1 < z < 5$. In the future, larger samples across a broad luminosity range and a wide redshift range are required to draw a firm conclusion about the redshift dependence. 

\subsection{Galaxy star formation property dependence of the fundamental plane}
\label{sec:FPgal}
For the first time, we studied the galaxy star formation property (SFGs and QGs) dependence of the fundamental plane whose best-fit parameters and source numbers are summarized in Table \ref{tab:BHFPparameter}.  We adopted the UVJ selection criteria in \cite{Schreiber2015} to decide galaxy types (SFG or QG). Here, the UVJ magnitudes of our sample were derived from literature works
(GOODS-N: \citealt{Barro2019}; GOODS-S: \citealt{Straatman2016}; COSMOS: \citealt{Weaver2022}).
We know that radio AGNs hosted by SFGs and QGs have different cosmic evolutions for the AGN incident rate \citep[e.g.,][Paper I]{Janssen2012,Kondapally2022} and radio luminosity functions \citep[][Paper I]{Kondapally2022}.
In addition, SFGs and QGs have different gas contents that may perform different fueling mechanisms toward central SMBHs \citep[e.g.,][]{Kauffmann2009,Kondapally2022,Ni2023}.
Therefore, we want to know whether the galaxy star formation properties affect the fundamental plane.
For our radio-quiet AGN sample, the scatter, $\sigma_{\rm R}$, of the QG-hosting subset is significantly lower than that of the entire radio-quiet AGN sample and the SFG-hosting subset (see Table \ref{tab:BHFPparameter}), and the $f$-test comparing the $\chi^2$ and $dof$ of the entire radio-quiet AGN sample and the QG-hosting subset gives a $p$ value much lower than 0.05. We performed the same statistical tests comparing the entire radio-quiet AGNs and the SFG-hosting subset, but the statistical tests do not show significant difference between their fundamental planes. This result can be expected due to the fact that the entire radio-quiet AGN sample is dominated by the SFG-hosting populations (see Table \ref{tab:BHFPparameter}). In conclusion, the fundamental plane of the radio-quiet AGNs shows a dependence on the galaxy star formation properties, which means that the disc-jet connection of the radio-quiet AGNs may be related to the fueling mechanisms of their host galaxies.
However, for the radio-loud AGN sample, we do not find a statistically significant dependence of the fundamental plane on the galaxy star formation properties: 1) the scatter, $\sigma_{\rm R}$, is not improved after splitting the radio-loud AGN sample into the SFG- and QG-hosting subsets (see Table \ref{tab:BHFPparameter}); and 2) the $f$-test $p$ value is higher than 0.05. Therefore, the disc-jet connection of the radio-loud AGNs may not be affected by the fueling mechanisms of their host galaxies.
In the future, larger samples are required to make more detailed analysis and draw a solid conclusion.
In addition, nearly 82\% of radio-quiet AGNs are hosted by SFGs and 18\% of the radio-quiet AGNs are hosted by QGs (see Table \ref{tab:BHFPparameter} and Panel B of Fig. \ref{fig:LRLXdis}), while the fraction of the radio-loud AGNs hosted by SFGs and QGs are both around 50\% (see Table \ref{tab:BHFPparameter} and Panel C of Fig. \ref{fig:LRLXdis}).

\subsection{Central engine for radio-quiet and radio-loud active galactic nuclei}
\label{sec:engine}
As we mentioned in Section \ref{sec:FPstudies}, radio-quiet AGNs and radio-loud AGNs are found to follow different fundamental planes, which may correspond to different accretion physics and different origins of X-ray radiation for the central engine (see Fig. \ref{fig:RXRMCCslope}). In Fig. \ref{fig:cartoon}, we show a cartoon illustrating the central engines of radio-quiet AGNs and radio-loud AGNs based on the observational results.

According to the observational evidence from our work and literature works (see details in Section \ref{sec:FPEdd}), X-rays emissions of radio-quiet AGNs at $0.01 < \lambda_{\rm Edd} < 0.1$ seem to be produced by a combination of ADAF and optically thin synchrotron radiation from the jet, while at $0.1 < \lambda_{\rm Edd} < 1$ they are mainly produced by the synchrotron radiation from the jet (see Panel A in Fig. \ref{fig:RXRMCCproperties} or the cartoon illustration in Fig. \ref{fig:cartoon}). For radio-quiet AGNs at other $\lambda_{\rm Edd}$ ranges, more observational data are required to verify this trend in the future. \cite{Wang2022b} studied the ambient circumnuclear medium of six radio-quiet AGNs with $0.05 < \lambda_{\rm Edd} < 0.5$ based on the high-resolution X-ray analysis for warm absorber outflows. The density profile of the ambient circumnuclear medium shows that for these six radio-quiet AGNs, the accretion physics from 0.01 parsec to a few parsec \citep[corresponding to a physical scale from the broad-line region to the torus;][]{Wang2022a} are consistent with the standard thin disc model. One possible reason for the difference between this work and \cite{Wang2022b} is different physical scales. The fundamental plane describes the disc-jet or corona-jet connection in the innermost region of the accretion flow \citep{Merloni2003}, which reflects the accretion physics around several Schwarzschild radii \citep[e.g.,][]{Alston2020} (corresponding to about $10^{-4}$ parsec assuming a $10^8\ M_\odot$ black hole). Thus, the physical scale that we target in this work is much smaller than that in \cite{Wang2022b}.

The origins of X-ray emissions of radio-loud AGNs are consistent with a combination of ADAF and a synchrotron jet model at $\lambda_{\rm Edd} < 0.01$, agree with the synchrotron jet model at $0.01 < \lambda_{\rm Edd} < 0.1$, and follow the standard thin disc coupled with a synchrotron jet model at $\lambda_{\rm Edd} > 0.1$ (see Panel A in Fig. \ref{fig:RXRMCCproperties} or the cartoon illustration in Fig. \ref{fig:cartoon}).
The result for the high-$\lambda_{\rm Edd}$ radio-loud AGN is consistent with that in \cite{Zhong2023}, who found that the extremely powerful radio jets are a result of either the spectral transition from ADAF to a thin disc or a super-Eddington accretion.
 
\section{Summary and conclusions}
\label{sec:summary}

We investigated the fundamental plane of black hole activity based on a large radio AGN sample selected from the GOODS-N, GOODS-S, and COSMOS/UltraVISTA fields across $0.1 < z \leq 4$.
This radio AGN sample consists of 208 objects with available estimates for rest-frame 5 GHz radio luminosity ($L_{\rm R}$) and rest-frame 2--10 keV X-ray luminosity ($L_{\rm X}$), and a black hole mass ($M_{\rm BH}$) inferred from the stellar mass (Section \ref{sec:physicalpara}). 
We divided this radio AGN sample into 141 radio-quiet AGNs and 67 radio-loud AGNs (Section \ref{sec:RQRLsample}), and studied their fundamental planes separately (Section \ref{sec:fitness}). Further, we summarized the current studies about the fundamental plane (Section \ref{sec:FPstudies}). Finally, we explored the dependence of the fundamental plane on the Eddington ratio (Section \ref{sec:FPEdd}), redshift (Section \ref{sec:FPz}), and galaxy star formation properties (Section \ref{sec:FPgal}), and discussed the central engines of radio-quiet and radio-loud AGNs (Section \ref{sec:engine}). The main conclusions are shown as follows:
\begin{enumerate}
\item[(i)] The ratio of $L_{\rm R}$ to $L_{\rm X}$, also known as a tracer for radio loudness, shows a bimodal distribution that can be described well by two single Gaussian models (Fig. \ref{fig:LRLXdis}). The cross point between these two Gaussian components — that is, $\log (L_{\rm R}/L_{\rm X}) = -2.73$ — is defined as a radio-loudness threshold to divide radio-quiet and radio-loud AGN in this work.
\item[(ii)] Our radio-quiet AGNs have a larger X-ray luminosity, lower radio luminosity, and significantly larger Eddington ratio than our radio-loud AGNs.
\item[(iii)] Our radio-quiet and radio-loud AGNs show a significantly different fundamental plane: $\log L_{\rm R} = (0.47\pm 0.06)\log L_{\rm X}+(0.29\pm 0.06)\log M_{\rm BH}+17.06^{+2.45}_{-2.47}$ for the radio-quiet AGNs, and $\log L_{\rm R} = (0.74\pm 0.09)\log L_{\rm X}+(0.09\pm 0.09)\log M_{\rm BH}+8.35^{+3.61}_{-3.60}$ for the radio-loud AGNs (Fig. \ref{fig:BHFPfit} and Table \ref{tab:fitness}). Our radio-quiet AGNs can be explained by an ADAF model coupled with a synchrotron jet model, while our radio-loud AGNs mainly agree with a synchrotron-emitting jet model (Fig. \ref{fig:RXRMCCslope}). 
\item[(iv)]  For both radio-quiet and radio-loud AGNs, the fundamental plane shows a significant dependence on the Eddington ratio (Panel A in Fig. \ref{fig:RXRMCCproperties}), but no significant dependence on redshift (Panels B in Fig. \ref{fig:RXRMCCproperties}, respectively). The fundamental plane of radio-quiet AGNs show a dependence on the galaxy star formation properties (SFGs and QGs), while for radio-loud AGNs this dependence disappears.
\item[(v)] Radio-quiet and radio-loud AGNs at different accretion states have different fundamental planes (Fig. \ref{fig:RXRMCCslope}) corresponding to different accretion physics and different origins of X-ray emission (Fig. \ref{fig:cartoon}).
X-ray emissions of radio-quiet AGNs at $0.01 < \lambda_{\rm Edd} < 0.1$ are produced by a combination of ADAF and a synchrotron jet model, while at $0.1 < \lambda_{\rm Edd} < 1$ they mainly follow the synchrotron jet model (Fig. \ref{fig:cartoon}). The origins of X-ray emission of radio-loud AGNs are consistent with a combination of ADAF and a synchrotron jet model at $\lambda_{\rm Edd} < 0.01$, agree with the synchrotron jet model at $0.01 < \lambda_{\rm Edd} < 0.1$, and follow the standard thin disc coupled with the jet model at $\lambda_{\rm Edd} > 0.1$ (Fig. \ref{fig:cartoon}).
\end{enumerate}

 \begin{acknowledgements}
 We thank the anonymous referee for the constructive comments that greatly improved this paper.
 We thank Stefano Marchesi and Giorgio Lanzuisi for kind help to access the COSMOS-Legacy survey data.
 This work is supported by the National Natural Science Foundation of China (Project No. 12173017 and Key Project No. 12141301). LCH was supported by the National Science Foundation of China (11991052, 12011540375, 12233001), the National Key R\&D Program of China (2022YFF0503401), and the China Manned Space Project (CMS-CSST-2021-A04, CMS-CSST-2021-A06).

 \end{acknowledgements}

\bibliographystyle{aa}
\bibliography{ms.bib}

\end{CJK*}
\end{document}